\documentstyle[12pt]{article}
\def\k{{\rm {\bf k}}}
\def\p{{\rm {\bf p}}}

\def\hp{{\rm {\hat p}}}
\def\hk{{\rm {\hat k}}}
\begin{document}

\noindent Yukawa Institute Kyoto

\hfill YITP/K-1071

\hfill April 1994

\vspace{1.5cm}

\begin{center}
{\bf INFRARED PROPERTIES OF THE HOT GAUGE THEORY
AFTER SYMMETRY BREAKING}

\vspace{1.5cm}
{\bf O.K.Kalashnikov}
\footnote{Permanent address: Department of
Theoretical Physics, P.N.Lebedev Physical Institute, Russian
Academy of Sciences, 117924 Moscow, Russia. E-mail address:
kalash@td.fian.free.net}

Yukawa Institute for Theoretical Physics

Kyoto University,

Kyoto 606-01, Japan

\vspace{2.5cm}

{\bf ABSTRACT}

\end{center}

It is shown that the fictitious infrared pole is eliminated
from the hot gauge theory which acquires a new vacuum after
the global gauge symmetry spontaneously breaking. The
nonzero W-condensate is generated and leads to the screening
of the chromomagnetic forces through the scenario with the
standard magnetic mass.

\newpage

\section {Introduction}

The strong infrared divergences which usually accompany the
multi-loop perturbative calculations with a hot non-Abelian
gauge theory are an intrinsic property of SU(N)-gluodynamics
and determine many of its peculiarities. Unfortunately, their
essence remains unclear, although the infrared properties  of
Yang-Mills theory at $T\ne 0$ have been intensively studied
during recent years. At $T\ne 0$ the infrared problem in QCD
is more pronounced since, unlike QFT, the leading infrared
diverginces are power-like and most of the Green functions
calculated perturbatively demonstrate the fictitious infrared
pole when $p_4=0$ and $|\p|\sim g^2T$. This singularity
(originally found in papers [1,2]) is strongly sensitive to
the gauge chosen although cannot be removed through its
redefinition and remains prartically the same when the
radiative corrections are taken into account.

In present paper we study the infrared properties of the hot
SU(2)-model which acquires a new vacuum with the nonzero
W-condensate after the global gauge symmetry spontaneously
breaking. Within this scenario the infrared fictitious pole
being a signal of the infrared instability for the unbroken
gauge theory is eliminated at the beginning and the standard
magnetic mass arises already on the one-loop level. The broken
gauge theory demonstrates the correct infrared properties but
unfortunately all its infrared limits are strongly
gauge-dependent at least within a perturbative expansion. This
dependence seems to be not occasional and its apperance should
be carefully investegated although it is necessary to remember
(see e.g. Ref.[3] for details) that all the standard
nonperturbative infrared limits are gauge-dependent as well.

\section {General formalism and definition}

Our formalism is built by using the standard Green function
technique at $T\neq 0$ in the background
gauge with an arbitrary parameter $\xi$. The effective action
$W({\bar A})$ is defined as a functional integral over periodic
gauge and ghost fields
\setcounter{equation}{0}
\begin{eqnarray}
 \exp[-W({\bar A})V/T]=
N\int{\cal D}A{\cal D}C{\cal D}{\bar C}
\exp{\left\{+\int\limits_0^{\beta}d\tau \int d^3x
({\cal L}+J_{\mu}^aV_{\mu}^a)
\right\}}
\end{eqnarray}
where N is a temperature independent normalization factor (here
$\beta=1/T$), $V$ is a space volume; $V_{\mu}^a$, $C$ an
${\bar C}$ are the quantum gauge fields and ghosts,
respectively; and $J_{\mu}^a$ is an external source.

The quantum Lagrangian in the background gauge has the
standard form
\begin{eqnarray}
{\cal L}=-\frac{1}{4}(G_{\mu\nu}^a)^2+\frac{1}{2\xi}
[({\bar {\cal D}}_{\mu}V_{\mu})^a]^2 +
{\bar C}{\bar {\cal D}}_{\mu}\nabla_{\mu}C
\end{eqnarray}
where the gauge fields $ A_{\mu}^a $ are decomposed in the
quantum part $ V_{\mu}^a $ and the classical constant one
$ \bar {A}_{\mu}^a $ (here $ A_{\mu}^a=V_{\mu}^a+\bar {A}_{
\mu}^a $ ). The gauge fields strength tensor $G_{\mu\nu}^a$ is
determined through a new covariant derivative
${\bar {\cal D}}_{\mu}^{ab}=\partial_{\mu}{\delta}^{ab}+
gf^{acb}\bar {A}_{\mu}^c$ in the standard manner and the same
derivative defines the gauge fixing term. The term with ghost
fields contain two different derivatives where (unlike
${\bar {\cal D}}_{\mu}$) $\nabla_{\mu}$ is the standard
covariant derivative with the initial $A_{\mu}^a$-field.

To simplify our calculations we consider only the SU(2)-gauge
group and the external classical field is chosen to be
\begin{eqnarray}
\bar {A}_{\mu}^a=\delta_{\mu4}\delta^{a3}A^{ext}=
\delta_{\mu4}\delta^{a3}\frac{\pi T}{g} x
\end{eqnarray}
where $x$ is a new variable. Here $T$ is temperature and $g$
is the standard coupling constant. The parameter $ \xi $ fixes
the internal gauge and is very essential for what follows.

The nonzero external field breaks the global gauge symmetry
and SU(2)-fields acquire the different properties. In our case
there are two gauge sectors (transversal and longitudinal)
which involve the external  ${\bar A}_{\mu}^a$-field in the
different manner. The most pronounced changes take place
whithin the transversal gauge sector (for $V_\mu^\pm$-fields)
where the polarization tensor is not transversal already
on the one-loop level
\begin{eqnarray}
&&\Pi_{\mu\nu}^{\perp}(\p,\hp_4)=A^{\perp}(\p,\hp_4)
[\delta_{\mu\nu}
-\frac{\hp_\mu \hp_\nu}{\hp^2}]\nonumber\\
&& +B^{\perp}(\p,\hp_4)
[\frac{\hp_\mu \hp_\nu}{\hp^2}-\frac{u_\mu\hp_\nu+u_\nu\hp_\mu}
{({\bf u}\hp)}+\frac{u_\mu u_\nu}{({\bf u}\hp)^2}\hp^2 ]
+\frac{u_\mu u_\nu}{({\bf u}\hp)^2}D
\end{eqnarray}
and all functions depend on a new $\hp_4=(p_4+{\bar A}/g)$-
variable. Here $D=2ig^2\rho$ where $\rho$ is an external density.
Below Eq.(4) will be confirmed by the exact Slavnov-Taylor
identities and the explicit calculations made on the one-loop
level. Within the multi-loop calculations Eq.(4) should be more
complicated but its exact form is not found.

\section {A nontrivial vacuum of the SU(2)-model}

The effective action for the SU(2)-model (including the
two-loop graphs) has been calculated by many authors (see e.g.
Refs.[4,5] for the case when the $\xi$-parameter is arbitrary)
and has a rather simple form
\begin{eqnarray}
&&W(x)/T^4=W^{(1)}(x)/T^4+W^{(2)}(x)/T^4,\nonumber\\
&&W^{(1)}(x)/T^4=\frac{2}{3}\pi^2[B_4(0)+2B_4(\frac{x}{2})],
\nonumber\\
&&W^{(2)}(x)/T^4=\frac{g^2}{2}[B_2^2(\frac{x}{2})+
2B_2(\frac{x}{2})B_2(0)]+\frac{2}{3}g^2(1-
\xi)B_3(\frac{x}{2})B_1(\frac{x}{2}),\nonumber\\
\end{eqnarray}
where $ B_n(z) $ are the modified Bernoulli polynomials
\begin{eqnarray}
&&B_1(z)=z-\epsilon(z)/2\,\,\,,\qquad
B_3(z)=z^3-3\epsilon(z)z^2/2+z/2,\nonumber \\
&&B_2(z)=z^2-|z|+1/6\,,\quad B_4(z)=z^4-2|z|^3+z^2-1/30
\nonumber\\
\end{eqnarray}
with $\epsilon(z)=z/|z|$. Here we consider that $\epsilon
(0)=0$ to make (5) in accordance with the direct calculations.
Of couse, the effective action $W(x)$ is a gauge-dependent
quantity but the gauge covariance within Eq.(5) has been
demonstrated in detail [5].

To determine a new vacuum (or a possibility to have the nozero
W-condensate) one should find the nontrivial solution of the
standard extemum equation which is obtained within Eq.(5) to be
\begin{eqnarray}
&&B_3(\frac{{\bar x}}{2})+ \frac{3g^2}{8\pi^2}B_1(\frac{{\bar
x}}{2})[B_2(\frac{{\bar x}}{2})+ B_2(0)]\nonumber\\
&& +\frac{g^2}{8\pi^2}(1-\xi)[3B_2(\frac{{\bar x}}{2})
B_1(\frac{{\bar x}}{2})+B_3(\frac{{\bar x}}{2})]=0
\end{eqnarray}
and this equation (the same as $W(x)$) is $\xi$-dependent
explicitly. In our case (besides the previously known point
${\bar x}_2=1$) Eq.(7) has a new solution [6]
\begin{eqnarray}
{\bar x}_{1,3}=1\pm\left\{1-\frac{
\frac{g^2}{2\pi^2}[1+(1-\xi)/2]}
{1+\frac{3g^2}{8\pi^2}[1+\frac{4}{3}(1-\xi)]}
\right\}^{1/2}
\end{eqnarray}
which being nontrivial is a good candidate for a new vacuum.
For a small g these points are proportional to
the $g^2$-terms
\begin{eqnarray}
\bar{x}_{1,3}=1\pm\left[1-\frac{g^2}{4{\pi}^2}(\frac{3
-\xi}{2})\right]
\end{eqnarray}
and we use this solution to calculate a new effective action
after symmetry breaking.

Unfortunately we do not know the complete effective action up to
$g^4$-order terms and therefore only a part of $g^4$-corrections
are acceptable when the thermodynamical potential
is calculated. Thus being substituted to the effective action
(5) the new extremum points generate the gauge-dependent
$g^4$-corrections [7]
\begin{eqnarray}
\Omega/T^4
&=&\Omega^{(1)}/T^4+\Omega^{(2)}/T^4
=2\pi^2B_4(0)+\frac{3g^2}{2}B_2^2(0)\nonumber\\
&+&\frac{g^4}{48\pi^2}(\frac{3-\xi}{2})^2-\frac{g^4}{24\pi^2}
|\frac{3-\xi}{2}|(\frac{3-\xi}{2})\nonumber\\
\end{eqnarray}
which are (of course) only a part of the $g^4$-thermodynamical
potential but very important for what follows. Indeed with the
aid of Eq.(10) we can find perturbatively the difference between
two potentials which determine the initial and broken
SU(2)-theory. Our result has the form
\begin{eqnarray}
\Delta\Omega/T^4&=&(\Omega^/T^4)_{x\ne 0}
-(\Omega^/T^4)_{x=0}\nonumber\\
&=&\frac{g^4}{48\pi^2}(\frac{3-\xi}{2})^2-\frac{g^4}{24\pi^2}
|\frac{3-\xi}{2}|(\frac{3-\xi}{2})
\end{eqnarray}
and we believe that this expession
is exact up to $g_4$-order terms since
the unknown part of the $g^4$-effective action (being
treated by keeping calculational accuracy with the aid of
the trivial extremum points which are not depend on $g^2$)
is dropped out from Eq.(11). So, in accordance with Eq.(11),
we can conclude that for any $\xi<3$ the new vacuum formed by
the nonzero W-condensate is preferable and can be considered
as a possible physical state despite its crucial dependence
from $\xi$. Below we accept this scenario and study
the new infrared properties acquired by the broken theory.

\section {Infrared properties after the global gauge
symmetry breaking}

A polarization tensor for the broken SU(2)-group (when $x\ne 0$)
has two parts $\Pi^{||}(\p,p_4)$ and $\Pi^{\perp}(\p,\hp_4)$
which can be completely separated within the
$g^2$-approximation. Their calculation is standard and exploits
the usual temperature Green functions technique in the imaginary
time space. To simplify what follows most of details are omitted
and below the infrared limits of the $\Pi_{44}(\p,p_4)$-
components are presented only by their leading terms.

\subsection{SU(2)-transversal sector}

This sector contains two gauge fields $V_\mu^\pm =\frac{1}
{\sqrt 2}(V_\mu^1 \mp iV_\mu ^2)$ which do not commutate with
the external ${\bar A}_\mu^a$-field. The ghost ones are included
as usual. The inverse gluon propagator is defined through
$\Pi_{\mu\nu}^{\perp}(\p,\hp_4)$-tensor in the standard manner
\begin{eqnarray}
{\cal D}^{-1}(\p,\hp_4)={\cal D}_0^{-1}(\p,\hp_4)
+\Pi^{\perp}(\p,\hp_4)
\end{eqnarray}
but a new vacuum redefines the $p_4$-dependence of all
functions from the SU(2)-transversal sector. Here $\hp_4=
p_4+\mu$ where $\mu=g^2T(3-\xi)/8\pi$ as it was found above.
Unlike the case $\mu=0$, the polarization tensor
is not transversal and has a more complicated tensor structure
\begin{eqnarray}
&& \Pi_{ij}^{\perp}(\p,\hp_4)=\nonumber\\
&& =[\delta_{ij}-\frac{p_ip_j}{\p^2}]A^{\perp}(\p,\hp_4) +
\frac{p_i p_j}{\p^2}\frac{\hp_4^2}{\p^2}
[\Pi_{44}^{\perp}(\p,\hp_4)-2i{g^2}
{\rho}/{\hp_4}],\nonumber\\
&& \Pi_{i4}^{\perp}(\p,\hp_4)=-\frac{\p_i \hp_4}{\p^2}
[\Pi_{44}^{\perp}(\p,\hp_4)-2i{g^2}{\rho}/{\hp_4}]
\end{eqnarray}
but only two scalar functions $\Pi_{44}^{\perp}
(\p,\hp_4)$ and $A^{\perp}(\p,\hp_4)$ should be calculated
analytically. The decomposition (13) is confirmed by
the exact Slavnov-Taylor identity which (unlike the case
$\mu=0$ ) also acquires the additional term with an external
sourse [8]. For the one-loop calculations in the Feynman gauge
(where $\xi=1$) this identity can be simplified as follows
\begin{eqnarray}
{\hp}_\mu\Pi_{\mu\nu}^{\perp}(\p,\hp_4)=-gJ_\nu^3
\end{eqnarray}
where $J_\nu^3=-2ig\rho u_\nu$ to be on the same level. The
spontaneously generated density $\rho$ has the form
\begin{eqnarray}
\rho=\int\frac{d^3\k}{(2\pi)^3}(n^+-n^-)\,,\qquad
n^{\pm}=\{exp[\beta(\k\pm i\mu)]-1\}^{-1}
\end{eqnarray}
and demonstrates that Eq.(12) is in accordance with Eq.(14).

After all one-loop diagrams being gathered (there are the four
different ones) the $\Pi_{\mu\nu}^{\perp}(\p,\hp_4)$-tensor is
found to be
\begin{eqnarray}
&& -\Pi_{ij}^{\perp}(\p,\hp_4)=\frac{1}{\beta}\sum_{k_4}\int
\frac{d^3\k}{(2\pi)^3}\left\{\frac{g^2}{k^2(k+\hp)^2}
[\delta_{\mu\nu}(2k^2+5\hp^2+2k\hp)
\right.\nonumber\\&& \left.+8k_\mu k_\nu-
2\hp_\mu\hp_\nu+4(k_\mu\hp_\nu+k_\nu\hp_\mu)]-3\delta_{\mu\nu}
g^2\left(\frac{1}{k^2}+\frac{1}{\hk^2}\right)\right\}
\end{eqnarray}
and its infrared limits will be investigated below. We calculate
these limits in the standard infrared manner (when the whole sum
over $k_4$ is replaced by one term with $k_4=0$) and discuss
them in accordance with another calculations.

The $\Pi_{44}^{\perp}(0,\mu)$-limit will be calculated at first
to check Eq.(13) and all other formulae
found above. Moreover there is another reason to calculate
the $\Pi_{44}^{\perp}(0,\mu)$-limit carefully because namely
this limit determines the nozero extra density which
spontaneously generates after the global gauge symmetry
breaking. The $\Pi_{44}^{\perp}(0,\mu)$-limit has a rather
simple form
\begin{eqnarray}
\Pi_{44}^{\perp}(0,\mu)=
\frac{1}{\beta}\sum_{k_4}\int
\frac{d^3\k}{(2\pi)^3}\left\{\frac{g^2}{k^2\hk^2}
[2\mu^2+8\k^2]-
2g^2\left(\frac{1}{k^2}+\frac{1}{\hk^2}\right)\right\}
\end{eqnarray}
and can be found directly from Eq.(16) with the aid of the
standard algebra. This limit is easily calculated if we use
the next integral
\begin{eqnarray}
&& \frac{1}{\beta}\sum_{k_4}
\frac{1}{(k_4^2+\k^2)[(k_4+\mu)^2+\k^2]}\nonumber\\
&& =-\frac{1}{|\k|}\frac{1}{\mu^2+4\k^2}\left(n+\frac{n^+
+n^-}{2}\right)+\frac{i\mu(n^+-n^-)}{\mu^2(\mu^2+4\k^2)}
\end{eqnarray}
and another ones to reproduce the exact summation over $k_4$.
The final result has the form
\begin{eqnarray}
\Pi_{44}^{\perp}(0,\mu)=\frac{ig^2}{\mu}\int\frac{d^3\k}
{(2\pi)^3}(n^+-n^-)
\end{eqnarray}
and demonstrates a selfconsistency all calculations made since
the same expressions can be found directly from Eq.(13). Of
course, Eq.(19) correctly reproduces the known $\Pi_{44}(0)$-
limit if $\mu$ goes to zero. Here one should exploit the
following formula
\begin{eqnarray}
(n^+-n^-)_{\mu\rightarrow 0}=-\frac{(2i\mu\beta)
exp(\beta |\k|)}{[exp(\beta |\k|)-1]^2}
\end{eqnarray}
and calculate exactly the remaining integral. The result is
\begin{eqnarray}
\Pi_{44}^{\perp}(0,\mu=0)=\frac{2g^2T^2}{3}
\end{eqnarray}
as it takes place for the unbroken SU(2)-theory (see e.g. Ref.
[9]).

Now our goal is to calculate the infrared limit for the function
$A^{\perp}(\p,\hp_4)$ which defines the transversal ${\cal D}_
{ij}^{\perp}(\p,\hp_4)$-propagator. Namely this function has a
"wrong" behaviour which leads to the fictitious pole of the
unbroken SU(2)-theory.
Here the $\Pi_{ij}^{\perp}(\p,\hp_4)$-tensor
is not calculated completely but the function $A^{\perp}
(\p,\hp_4)$ can be found through the special representation
\begin{eqnarray}
A^{\perp}(\p,\hp_4)=\frac{1}{2}\sum_i
\left[\Pi_{ii}^{\perp}(\p,\hp_4)
+\frac{\hp_4 p_i}{\p_2}\Pi_{i4}^{\perp}(\p,\hp_4)\right]
\end{eqnarray}
which takes place in accordance with Eq.(13). There are two
terms in Eq.(22) which should be calculated independly but in
the same manner. At first we calculate the $p_i\Pi_{i4}^{\perp}
(\p,\hp_4)$-term which is found to be
\begin{eqnarray}
p_i\Pi_{i4}^{\perp}(\p,\hp_4)=\frac{g^2}{\beta}\sum_{k_4}
\int\frac{d^3\k}{(2\pi)^3}\frac{
2\hp_4(\p^2-2\p\k)-4k_4(\p^2+2\p\k)}
{k^2(k+\hp)^2}
\end{eqnarray}
if Eq.(16) has been used. Performing calculation in the
infrared manner we simplify Eq.(23) as follows
\begin{eqnarray}
p_i\Pi_{i4}^{\perp}(\p,\mu)=\frac{2g^2\mu}{\beta}
\int\frac{d^3\k}{(2\pi)^3}\frac{\p^2-2\p\k}
{\k^2[\mu^2+(\k+\p)^2]}
\end{eqnarray}
and then exactly calculate the remaining integral.
The result has the form
\begin{eqnarray}
p_i\Pi_{i4}^{\perp}(\p,\mu)=\frac{g^2}{\beta}\left[-\frac{\mu^2}
{2\pi}+\left(\frac{\mu^3}{2\pi|\p|}+\frac{\mu|\p|}{\pi}\right)
Arctan[\frac{|\p|}{\mu}]\right]
\end{eqnarray}
and demonstrates that when $\mu\ne 0$ its infrared limit is
completely different from the case $\mu=0$. The same behaviour
takes place for the function $A^{\perp}(\p,\mu)$ which
is found to be
\begin{eqnarray}
A^{\perp}(\p,\mu)=-\frac{g^2}{\beta}\left[\frac{\mu}{4\pi}
+\frac{\mu^3}{4\pi\p^2}+\left(\frac{\mu^2}{2\pi|\p|}
+\frac{3\p|}{4\pi}-\frac{\mu^4}{4\pi|\p|^3}\right)
Arctan[\frac{|\p|}{\mu}]\right]
\end{eqnarray}
in accordance with equation Eq.(22). If $\mu=0$ we have the
standard expression
\begin{eqnarray}
A^{\perp}(\p,0)=-\frac{3g^2|\p|}{8\beta}
\end{eqnarray}
but this is not the case when $\mu\ne 0$. For the broken SU(2)-
theory the infrared $A^{\perp}(\p,\mu)$-limit has the form
\begin{eqnarray}
A^{\perp}(\p,\mu)_{|\p|\rightarrow 0}=-\frac{g^2}{\beta}
\left[\frac{3\mu}{4\pi}+\frac{3\p^2}{4\pi\mu}\right]
\end{eqnarray}
and one can see that its infrared behaviour crucially changes.
The fictitious infrared pole disappears
\begin{eqnarray}
\left[\hp^2+A^{\perp}(\p,\hp_4)\right]_{p_4=0}=
-\left[\frac{g^4T^2(9-\xi^2)}{64\pi^2}+
\frac{3+\xi}{3-\xi}\p^2\right]
\end{eqnarray}
and after the global gauge symmetry spontaneously breaking
the gauge fields acquire the correct infrared behaviour.
Here we use that $\mu=g^2T(3-\xi)/8\pi$ as it was obtained above.
Of course, the found behaviour is very sensitive to the chosen
gauge but remains the same for all $\xi <3$. For $\xi > 3$
the situation changes however for the same $\xi$ the sign in
Eq.(11) also changes and the nontrivial vacuum will be
metastable.

\subsection{SU(2)-longitudinal sector}

This sector contains one gauge field $V_\mu^3$ which commutates
with external ${\bar A}_\mu^a$-field and the appropriate ghost
one. The inverse gluon propagator is defined through the
$\Pi_{\mu\nu}^{||}(\p,p_4)$-tensor in the standard manner
and a new vacuum keeps the usual $p_4$-dependence of all
functions from the SU(2)-longitudinal sector. Now the
polarization tensor is transversal and has the form
\begin{eqnarray}
&& -\Pi_{ij}^{||}(\p,p_4)=\frac{1}{\beta}\sum_{k_4}\int
\frac{d^3\k}{(2\pi)^3}\left\{\frac{g^2}{\hk^2(\hk+p)^2}
[\delta_{\mu\nu}(2\hk^2+5p^2+2\hk p)
\right.\nonumber\\&& \left.
+8\hk_\mu \hk_\nu-
2p_\mu p_\nu+4(\hk_\mu p_\nu+\hk_\nu p_\mu)]-
\frac{6\delta_{\mu\nu} g^2}{\hk^2}\right\}
\end{eqnarray}
where all abreviations are the same as previously. All program
performed above is acceptable within Eq.(30) but we explicitly
find only the $A^{||}(\p,0)$-limit since the latter is the main
point of our conclusion. This limit is easily calculated
through the standard condition
\begin{eqnarray}
A^{||}(\p,0)=\frac{1}{2}\sum_i\Pi_{ii}^{||}(\p,0)
\end{eqnarray}
which results from Eq.(13) for the case $\rho=0$.
After treating Eq.(30) in the infrared manner the
$A^{||}(\p,0)$-limit is found to be
\begin{eqnarray}
A^{||}(\p,0)=\frac{g^2}{2\beta}\int
\frac{d^3\k}{(2\pi)^3}\left\{\frac{6\mu^2+13\p^2+14(\k^2+\k\p)}
{(\mu^2+\k^2)[\mu^2+(\p+\k)^2]}
-\frac{18}{\mu^2+\k^2}\right\}
\end{eqnarray}
and below this integral is exactly calculated. Our result has
the form
\begin{eqnarray}
A^{||}(\p,0)=-\frac{g^2}{\beta}\left[\frac{\mu}{2\pi}+
\left(\frac{3|\p|}{4\pi}-\frac{\mu^2}{\pi|\p|}\right)
Arctan[\frac{|\p|}{2\mu}]\right]
\end{eqnarray}
which is similar Eq.(36) and demonstrates the same properties.
The $A^{||}(\p,0)$-limit changes to be
\begin{eqnarray}
A^{||}(\p,0)_{|\p|\rightarrow 0}=-\frac{3g^2\p^2}
{8\pi\beta\mu}
\end{eqnarray}
and acquires the normal infrared behaviour for small
$\p^2$. The correspondent pole structure also changes
\begin{eqnarray}
\left[p^2+A^{||}(\p,p_4)\right]_{p_4=0}=
-\frac{\xi}{3-\xi}\p^2
\end{eqnarray}
but essentially depends on the gauge in the same manner
as previously.

\section {Conclusion}

The results obtained above open a new possibility to eliminate
the infrared fictitious pole from the hot gauge theory which
acquires a new vacuum after the global gauge symmetry
spontaneously breaking. The broken gauge theory has at the
beginning a correct infrared behaviour of all Green
functions calculated perturbatively and this scenario is
simpler then any nonperturbative schemes known previously.
Here we confirm  that the nonzero magnetic mass seems to be
a inherent attribute of the hot gauge theory and now it arises
already on the one-loop level to be
\begin{eqnarray}
m_{M}^2=\frac{g^4T^2(9-\xi^2)}{64\pi^2}
\end{eqnarray}
This quantity is approximately the same as the magnetic mass
found previously (see e.g. Ref.[3]) but it is essential
that now the standard perturbative expansion is more reliable
and selfconsistent. Moreover this scenario has a chance to be
useful for the GUT models since the spontaneously generated
density $\rho$ can be connected with the baryon charge
to explain, for example, the problem with the baryonic number
generation. Of course, we hope that the multi-loops
corrections (especially when the fermions ones are taken
into account) will not spoil the presented scenario (see e.g.
Refs.{10]) and its gauge dependence will be eliminated.
However it may be not always the case on the $g^2T$-scale where
the gauge invariance seems to be questionable but the gauge
covariance should be kept throughout any calculations
(particularly it concerns the nonperturbative ones).

\section {Acknowledgements}

I would like to thanks all the colleagues from the Yukawa
Institute for the useful discussions and the kind hospitality.
Especially I am very obliged to Professor Y. Nagaoka for
inviting me to the Yukawa Institute where this investigation
has been performed.

\newpage

\section {References}

\renewcommand{\labelenumi}{\arabic{enumi}.)}
\begin{enumerate}

\item{ O.K.Kalashnikov and V.V.Klimov. Sov. J. Nucl. Phys.
{\bf 33} (1981) 443; [ Yad. Fiz. {\bf 33} (1981) 848 ] }

\item{ R. Jackiw and S.Templeton. Phys. Rev.{\bf D23} (1981)
2291;\\ T.Appelquist and R.D.Pisarski. {\bf D23} (1981) 2305.}

\item{ O.K.Kalashnikov. Preprint BNL-46878, NUHEP-TH-91 \\
(October 1991); Phys. Lett. {\bf B279} (1992) 367. }

\item{ V.M.Belyaev. Phys. Lett. {\bf B254} (1991) 153.}

\item{ V.V.Skalozub. Mod. Phys. Lett. {\bf A7} (1992) 2895 }

\item{ O.K.Kalashnikov. Phys. Lett. {\bf B302} (1993) 453. }

\item{ O.K.Kalashnikov. JETP. Lett. {\bf 57} (1993) 773. }

\item{ O.K.Kalashnikov and  L.V. Razumov. Sov. Phys. Lebedev
Inst. Rep. {\bf N1} (1990) 71; [Sb. Kratk. Soobshch. Fiz.
{\bf N1} (1990) 50 ] }

\item{ O.K.Kalashnikov. Fortschr. Phys. {\bf 32} (1984) 525 }

\item{ A. Nakamura and K. Shiraishi. Acta Phys. Slov. {\bf 42}
(1992) 338. }

\end{enumerate}

\end{document}